# A Mutil-conditional Diffusion Transformer for Versatile Seismic Wave Generation

Longfei Duan[1,2†], Zicheng Zhang[3†], Lianqing Zhou[2*], Congying Han[3],

Lei Bai[4], Tiande Guo[3], and Cuiping Zhao[2*]

[1]Institute of Geophysics, China Earthquake Administration, 100081, Beijing, China

[2]Institute of Earthquake Forecasting, China Earthquake Administration, 100036, Beijing, China

[3]School of Mathematical Sciences, University of Chinese Academy of Sciences, 100049, Beijing, China

[4]Shanghai Artificial Intelligence Laboratory, 200032, Shanghai, China

## Abstract

Seismic wave generation creates labeled waveform datasets for source parameter inversion, subsurface analysis, and, notably, training artificial intelligence seismology models. Traditionally, seismic wave generation has been time-consuming, and artificial intelligence methods using Generative Adversarial Networks often struggle with authenticity and stability. This study presents the Seismic Wave Generator, a multi-conditional diffusion model with transformers. Diffusion models generate high-quality, diverse, and stable outputs with robust denoising capabilities. They offer superior theoretical foundations and greater control over the generation process compared to other models. Transformers excel in seismic wave processing by capturing long-range dependencies and spatial-temporal patterns, improving feature extraction and prediction accuracy compared to traditional models. To evaluate the realism of the generated waveforms, we trained downstream models on generated data and compared their performance with models trained on real seismic waveforms. The seismic phase-picking model trained on generative data achieved 99% recall and precision on real seismic waveforms. Furthermore, the magnitude estimation model reduced prediction bias from uneven training data. These findings suggest that diffusion-based generation models can address the challenge of limited regional labeled data and hold promise for bridging gaps in observational data in the future.

[†]These authors contributed equally to this work.

## 1. Introduction

Seismology is the science that studies seismic waves and what they tell us about the structure of the Earth and the physics of earthquakes (Shearer, 2019). Deriving structures and physical processes from observed waveforms is known as an inversion process, while forming seismic waves represents a forward modeling process. This modeling not only simulates the formation and propagation of seismic waves (Johnson, 1974; Wang, 1999), but is also widely applied in seismic studies, including source parameter inversion (Zhao and Helmberger, 1994; Zhu and Helmberger, 1996; Donner et al., 2020), artificial intelligence (AI) seismic models training (Araya-Polo et al., 2019; Zhang et al., 2020; Kuang et al., 2021), and other related applications (Tromp, 2019; Li et al., 2023). Especially, for training seismological AI models, due to the differing waveform characteristics across various stations and regions, challenges arise, such as reduced generalization of models trained in one region when applied to others, inability to generalize due to variations in velocity structure, or the issue of insufficient training data from a single region.

However, previous methods for generating seismic waves have either been time-consuming or produced waveforms with limited authenticity. Traditional methods, such as the spectral element method (Komatitsch and Tromp, 1999) and the finite difference method (Moczo et al., 2007), simulate seismic wave propagation based on elastic dynamic equations, requiring detailed analyses of the earthquake source and propagation path, which increases both complexity and time. Additionally, a limited understanding of the Earth's interior can significantly affect the authenticity of the generated waveforms. Furthermore, in AI seismology, reliance on artificial prior knowledge may compromise the integrity of training datasets (Wang and Wang, 2021). AI, especially deep learning, has recently demonstrated its power in seismology (Mousavi and Beroza, 2022, 2023; Anikiev et al., 2023) by improving research speed, accuracy (Wang et al., 2020; Zhou et al., 2021; Feng et al., 2024) and introducing new perspectives (Hulbert et al., 2019; Rouet-Leduc et al., 2019; Li, 2022). In particular, Generative Adversarial Networks (GANs) has already been used to synthesize seismic waveforms based on simple conditions (Meier et al., 2019; Wang et al., 2021; Esfahani et al., 2023; Chen et al., 2024). These models are less effective and controllable for complicated tasks like source parameters-based simulations, suffer from mode collapse issues that hinder capturing the diversity of seismic data distributions, and exhibit weak scalability on large-scale datasets (Chen et al., 2023).

Recently, diffusion models have emerged as a novel type of deep generative model, which has been widely used in tasks related to seismic waves (Jiang et al., 2023; Lan and Huang, 2024; Li et al., 2024; Shi et al., 2024; Wang et al., 2024; Wang et al., 2024). These models progressively add noise to data and then learn to reverse this process to generate new samples (see Figure. 1) (Yang et al., 2024). While diffusion models require more computational resources and time than methods like GANs, they deliver superior quality and mode coverage (Dhariwal and Nichol, 2021; Ho et al., 2021), indicated by high likelihood (Huang et al., 2021; Song et al., 2021; Kingma et al., 2023). This makes them a favored choice for high-fidelity image generation and restoration tasks (Chen et al., 2020; Kollovieh et al., 2023). The initial diffusion model, based on two Markov chains (Sohl-Dickstein et al., 2015), was later advanced with the development of the denoising diffusion probabilistic model (DDPM) and the introduction of the convolutional U-Net architecture (Ho et al., 2020). Recently, Peebles and Xie (2023) introduced Diffusion Transformers (DiT), which enhance the DDPM by substituting the U-Net with transformers. This advancement supports models such as Sora and Stable Diffusion 3, enabling them to generate realistic and imaginative scenes from text prompts.

In this paper, we introduce the Seismic Wave Generator (SWaG), demonstrating the versatile capabilities of diffusion models in seismic wave synthesis and analysis. Inspired by DiT, SWaG integrates a transformer architecture with multiple embedders to condition the generation using station ID, arrival times, and magnitude. We compare the generated and real waveforms and use them to train models for seismic phase picking and magnitude determination from scratch. These results confirm the effectiveness of large-scale pre-trained generative AI in seismology and provide valuable tools and insights for further research.

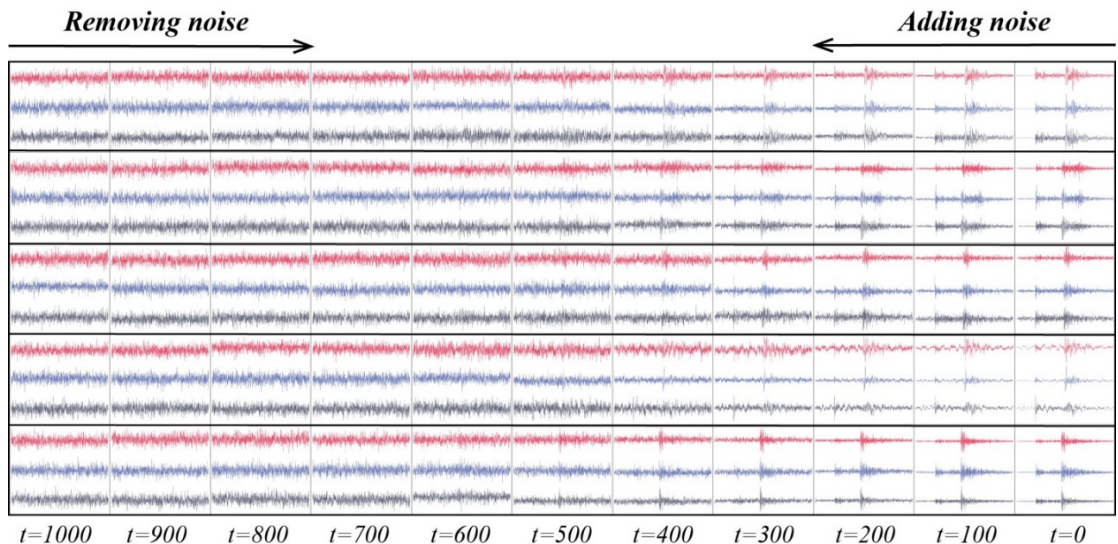

Figure 1: The process of adding and removing noise in diffusion models. From t=0 to t=1000 depicts the process of adding noise, while the reverse represents the process of removing noise.

## 2. Data and Methods

### 2.1 Diffusion Model

Diffusion models have emerged as a new class of deep generative models, which progressively corrupt data by adding noise and then learn to reverse this process to generate new samples. These models have surpassed the long-standing dominance of generative adversarial networks (GANs) and have set new benchmarks in various applications, including image synthesis, natural language processing, and molecule design. In this section, we provide a concise overview of the development and mathematical foundation of diffusion models.

The foundational work by Sohl-Dickstein et al. (2015) introduced the basic formulation of diffusion models, utilizing two Markov chains: a forward chain that perturbs data into noise, and a reverse chain that converts noise back to data. The forward chain is typically hand-designed to transform any data distribution into a simple prior distribution (e.g., standard Gaussian), while the reverse chain learns to reverse this process using transition kernels parameterized by deep neural networks. New data are generated by sampling random noise from the prior distribution and then applying ancestral sampling through the reverse Markov chain. Ho et al. (2020) advanced this concept with the denoising diffusion probabilistic model (DDPM) and introduced the convolutional U-Net architecture to the neural networks.

The diffusion model can be mathematically described as follows: Let $x_0$ be the real data and $x_T$ the randomly generated Gaussian noise data, where T represents the diffusion steps. The forward process adds Gaussian noise to $x_0$ iteratively:

$$x_t = \sqrt{\alpha_t} x_{t-1} + \sqrt{1 - \alpha_t} z \tag{1}$$

where $x_t$ is the noisy data at step $t$, $z \sim \mathcal{N}(0,1)$ is the noise from a standard normal distribution, and $\alpha_t$ is the coefficient at time step $t$. The reverse process denoises the data from $x_T$ to recover the original data:

$$x_{t-1} = \frac{1}{\sqrt{\alpha_t}} \left( x_t - \sqrt{1 - \alpha_t} z_\theta(x_t, t) \right) \tag{2}$$

where $z_\theta(x_t, t)$ is the noise at step $t$ predicted by the network.

The input to the network is the time $t$ and the labels, while the output is the predicted noise at step $t$. the model parameters are optimized using a simplified loss function:

$$\mathcal{L}_{simple} = \mathbb{E}[\|z - z_\theta(x_t, t)\|^2] \tag{3}$$

where $z \sim \mathcal{N}(0,1)$ is the noise from a standard normal distribution.

Numerous methods have been developed to further improve diffusion models, such as enhancing sampling techniques (e.g., classifier-free guidance), reformulating diffusion models, and using cascaded DDPM pipelines (Ho et al., 2021; Ho and Salimans, 2022; Gao et al., 2024). Meanwhile, a significant advantage of diffusion models is their scalability with increasing model size, training compute, and data, setting them apart from other generative models in applications (Croitoru et al., 2023). Inspired by the superior scaling properties of transformers compared to convolutional architectures, Peebles and Xie proposed Diffusion Transformers (DiT), which replace the commonly used U-Net backbone with a transformer to further enhance the scaling properties of DDPM (Peebles and Xie, 2023). Their comprehensive analysis of DiT found that models with higher computational complexity consistently performed better.

In addition to the empirically validated performance in various domains, we have recognized that diffusion models with a transformer architecture are particularly well-suited for seismic wave generation for two main reasons. First, the Markov chain, viewed as a stochastic differential equation, effectively simulates data with inherent noise—a critical feature of seismic waveforms (Weaver, 2011). Second, the transformer architecture has demonstrated considerable effectiveness in modeling sequential data (Vaswani et al., 2023), making it well-suited for capturing the complex temporal dependencies present in seismic waveforms (Mousavi et al., 2020).

## 2.2 Network Architecture

Our model inputs consist of noisy data $x_t$ at step $t$, with 3 dimensions (for the E, N, and Z components) and a length of 3000, matching the number of sampling points. The labels have 4 dimensions (station ID, P-wave arrival time, S-wave arrival time, and magnitude). The time step $t$ has a dimension and length of 1 (Figure 2).

To address the inefficiency of Transformers with long data sequences, we designed four convolutional neural network encoders to extract features from $x_t$. This results in a token of dimensions $768 \times 500$, serving as input for the DiT Blocks. Additionally, within this framework, we employ embedding layers to process both the labels and time vectors. The purpose of these embedding layers is to convert discrete data, which is inherently categorical and non-numeric, into continuous vector representations. This transformation is crucial because it enables the model to utilize the rich contextual information embedded within the categorical data. Furthermore, these embedding layers are trainable components of the overall model. This means that the embedding representations are continually refined to better capture the intrinsic structure and relationships within the data.

The model includes 28 DiT Blocks, each with a multi-head self-attention layer, Scale layers, and Layer Norm layers.

The self-attention layer captures global relationships within the waveform data. A multi-head cross-attention layer follows, integrating conditional inputs like labels and time vectors to enhance expressiveness. The Scale layers adjust input data using learned parameters, while the Layer Norm layers normalize the data for improved training stability and convergence.

The model training was conducted using four NVIDIA Tesla A800 GPU cards, each equipped with 80GB of memory. We set the batch size to 256 and used the Adam optimizer with a learning rate of 0.0001. The training process spanned 50 epochs and required approximately 40 hours to complete.

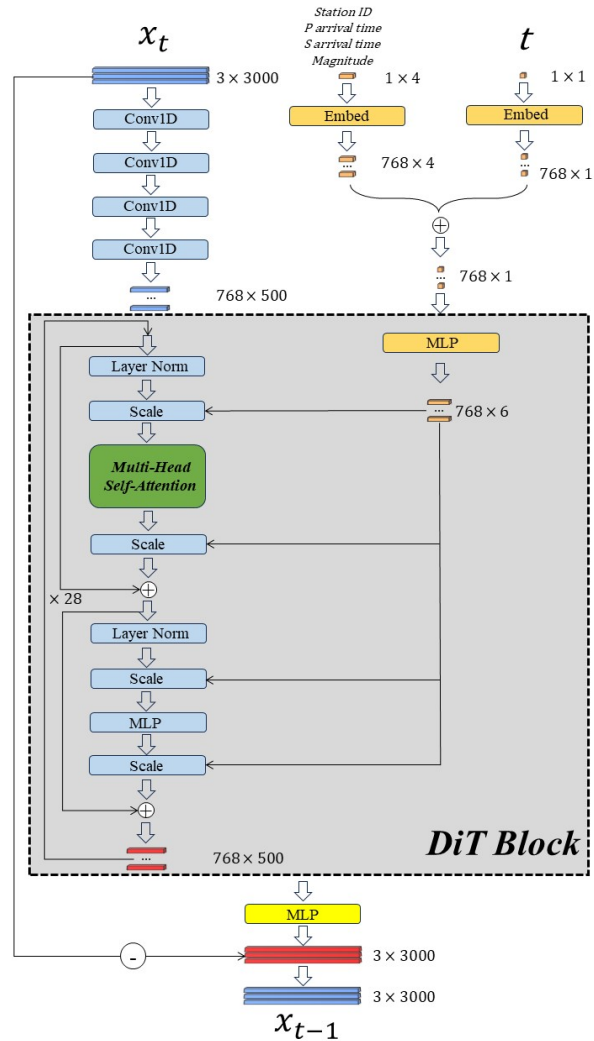

Figure 2. Network architecture of Seismic Wave Generator (SWaG).

### 2.3 Data and Labels

In this study, we utilized the STanford EArthquake Dataset (STEAD) to train the Seismic Wave Generator (SWaG).

STEAD is a high-quality, large-scale global earthquake waveform dataset comprising approximately 1.05 million three-component waveforms from local earthquakes. These waveforms were recorded by about 2,613 seismic stations worldwide, covering various instrument types, including broadband, high broadband, extremely short period, and short period instruments. Each waveform is 60 seconds long, sampled at 100 Hz. All waveforms were demeaned, detrended, and filtered. We retained only waveforms with an epicentral distance of less than 200 km, resulting in 100,588 three-component waveforms as train sets. These waveforms were downsampled to 50 Hz and normalized for training the SWaG. Additionally, station ID, P-wave arrival time, S-wave arrival time, and magnitude were included as input labels.

To validate generalization, we collected 8,000 three-component waveforms from a specific station with broadband instruments in the Sichuan-Yunnan region in China. The sampling rate at this station was 100 Hz, and none of the waveforms had been demeaned, detrended, or filtered. These waveforms were downsampled to 50 Hz and normalized. They were then randomly divided into transfer learning sets containing 5,000 waveforms and test sets containing 3,000 waveforms, each with P-wave arrival time, S-wave arrival time, and magnitude labels. The transfer learning sets are for the transfer learning of models such as SWaG, seismic phase picking models, and magnitude estimation models. The test sets evaluate how well models trained on generated waveforms predict real, unseen samples.

## 3. Comparison with Real Waveforms

We compared waveforms generated by SWaG with real waveforms, selecting from train sets and test sets, across signal segments, time domain, frequency domain, and time-frequency domain (Figure 3). Both seismic waveforms in the time domain and frequency domain are important because seismologists need them to extract crucial information for inversing seismic source parameters, rupture processes, and structures (Ishii et al., 2005; Zhao et al., 2008, 2011; Zhou et al., 2011; Avouac et al., 2015).

Comparing signal segments, the generated waveforms closely match the real ones with accurate P-wave and S-wave arrival times (Figure 3). The ratio of the S-wave and P-wave amplitudes in the horizontal components of the generated waveforms are significantly higher than in the vertical component, indicating that our model captures the three-component information. In the time domain, the decay length of the generated waveforms matches the real waveforms, confirming the effectiveness of the magnitude labels used in training. Notably, even without frequency information input, the frequency and time-frequency domain performances of the generated waveforms are very similar to the real waveforms.

The STEAD were preprocessed with demeaning, detrending, and filtering, while the transfer learning sets and test sets remained raw. Longitudinal comparisons show our model's strong generalization, quickly capturing specific information from regional waveforms. However, differences between the generated and real waveforms in time and frequency domains highlight the complexity of seismic waves under intricate source and propagation path conditions.

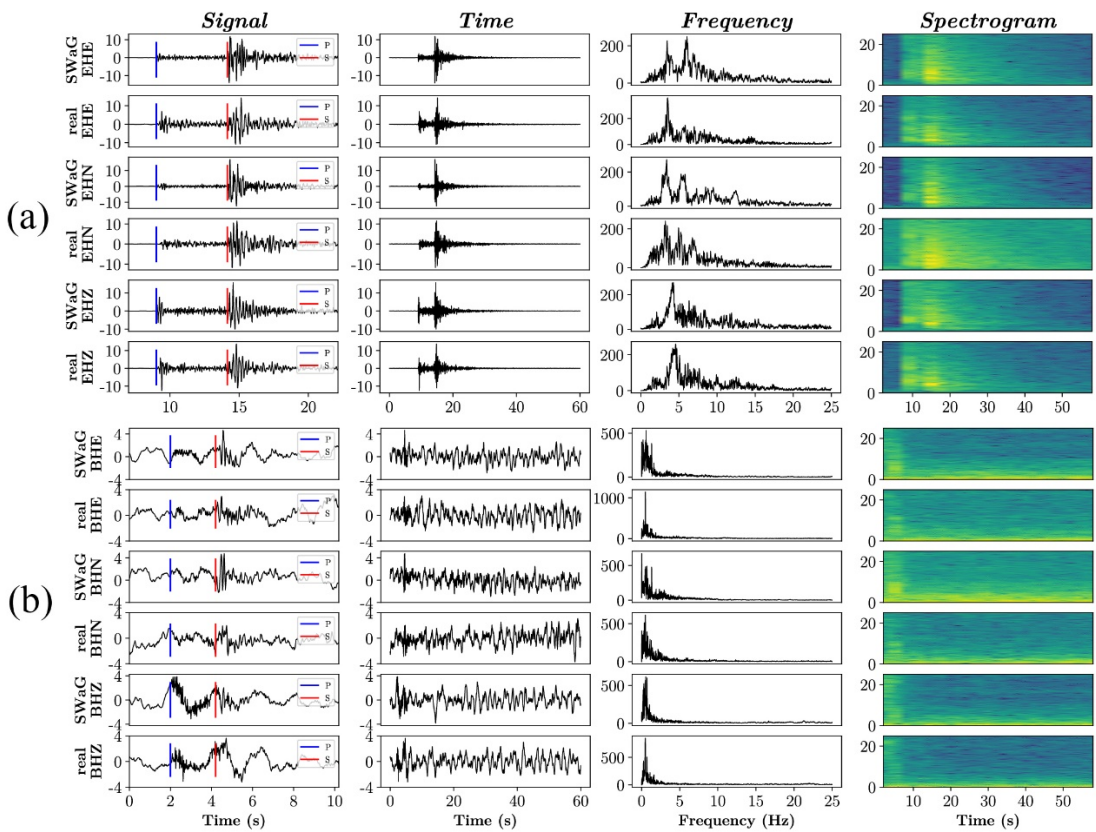

Figure 3. Comparison of generated and real waveforms in both STEAD and test datasets. (a) The comparison between the generated waveforms and the real waveforms with the same labels in the transfer learning set. These comparisons are made across the E, N, and Z components in the signal segments, time domain, frequency domain, and time-frequency domain. (b) The comparison between the generated waveforms and the real waveforms with the same labels in the test sets.

## 4. Downstream Model Training Using SWaG-Generated Waveforms

Assessing the quality of samples generated by generative models is inherently challenging, and no perfect method exists to evaluate their performance (Gm et al., 2020). The comparison in the section 3 is insufficient for demonstrating the quality of SWaG generation, as it is both singular and subjective. Additionally, statistical methods assessing the similarity between generated and real waveforms are not optimal, as real seismic waves are influenced by various factors, such as source, paths, receivers, and surface.

In this study, we decided not to directly evaluate the authenticity of each SWaG-generated sample. Instead, we trained two types of AI seismology models from scratch using only SWaG-generated waveforms to predict real and labeled waveforms, comparing their performance to a model trained on real waveforms. The first type of model is phase-picking model that extracts seismic signal from waveforms and identifies the arrival times of P- and S-phases. These models effectively distinguish between seismic signals and noise, capturing sudden changes in period and amplitude. The second model is designed for magnitude determination. Earthquake magnitude relates not only to the energy released but also to factors such as propagation, stations, and surface. These two experiments enable us to verify whether SWaG accurately captures the distribution of real waveforms and generates samples that align with the characteristics of seismic waves. They also confirm whether the generated waveforms satisfy the specified conditions of station ID, arrival time, and magnitude, thereby assessing the reliability of the labels. Furthermore, this demonstrates the potential for applying generated waveforms in various future domains, such as phase picking and magnitude estimation.

### 4.1 Downstream Model 1: Phase Piking Model

Seismic phase picking is a critical task in seismology for building earthquake catalogs (Ross et al., 2018; Zhu and Beroza, 2019; Mousavi et al., 2020; Hou et al., 2023; Zhao et al., 2024). However, models trained on standard datasets often show a marked decline in recall and precision when applied to regional data (Münchmeyer et al., 2022). In this study, we used Seisbench (Woollam et al., 2022) which is an open-source python toolbox for machine learning in seismology to develop six models from two universal phase-picking models: EQTransformer (Mousavi et al., 2020) and PhaseNet (Zhu & Beroza, 2019). Model Original-EQTransformer is the publicly available EQTransformer model trained on the STEAD dataset, while model Origianal-PhaseNet is the publicly available PhaseNet model trained on data from the Northern California Earthquake Data Center. Model Transfer-learning-EQTransformer is created by applying transfer learning to model Original-EQTransformer using the transfer learning sets, while model Transfer-learning-PhaseNet is developed through transfer learning applied to Origianal-PhaseNet. For model SWaG-EQTransformer, we first use the transfer learning sets for transfer learning on SWaG, followed by the random generation of 100,000 data labeled with P- and S-wave arrival times. On a server equipped with an A800 card, we generated one waveform every 0.2 seconds from random noise, whereas on a computer with an RTX3090 card, it took an average of 0.6 seconds per waveform. We then used this generated data to train the EQTransformer model from scratch with initialized parameters and got model SWaG-EQTransformer. Similarly, we developed model SWaG-PhaseNet. Finally, we employed these six models to make predictions on the test sets that was not used in training. We assessed the performance of each model using six metrics: recall, precision, F1-score, mean error, standard deviation of error, and mean absolute error (Table 1).

The model trained on generated waveforms outperforms those trained on real non-local waveforms and those using local data transfer learning, as assessed by precision, recall, and F1 score. This finding highlights four important points: 1. SWaG generates seismic waveforms that align with specified arrival labels and effectively captures the characteristics of seismic waveforms; 2. In cases of insufficient training data, fine-tuning SWaG can effectively capture local waveform features and yield more accurate predictions than fine-tuning EQTransformer and PhaseNet, especially EQTransformer and SWaG are trained on the same dataset; 3. The EQTransformer model is trained on a dataset of 1.2M waveforms, while the PhaseNet model uses approximately 780K waveforms. However, SWaG significantly outperforms models trained on only 100K waveforms, reducing training costs and demonstrating that a focused dataset can greatly enhance prediction accuracy; 4. When the training data is sufficiently abundant and evenly distributed, the differences between models will diminish. This suggests that region- or station-specific phase picking models, customized using waveforms generated by SWaG, can significantly overcome the challenges faced by previous phase picking models.

Additionally, we observed that the model trained on generated waveforms exhibits a slightly greater prediction bias than the model trained on real waveforms, particularly for S-waves. We attribute this to two potential reasons. First, the model trained on generated waveforms may have recalled more seismic phases with low signal-to-noise ratios, making predictions more challenging. Secondly, the SWaG using a 50Hz sampling rate, compared to the 100Hz of the original models, potentially doubling the time deviation for the same picking point.

Table 1. Comparison of picking results. μ and σ represent the mean and standard deviation of the error (real - predicted), while MAE represents the mean absolute error.

| | | EQTransformer | | | PhaseNet | | |
|---|---|---|---|---|---|---|---|
| Parameters | | 376,935 | | | 268,443 | | |
| Model | | Original-EQTransformer | Transfer-learning-EQTransformer | SWaG-EQTransformer | Original-PhaseNet | Transfer-learning-PhaseNet | SWaG-PhaseNet |
| Sampling Rate | | 100 Hz | 100 Hz | 50 Hz | 100 Hz | 100 Hz | 50 Hz |
| P | Recall | 0.31 | 0.49 | *0.99* | 0.98 | 0.88 | *0.98* |
| | Precision | 0.90 | 0.98 | *0.99* | 0.86 | 0.98 | *0.99* |
| | F1 | 0.46 | 0.65 | *0.99* | 0.92 | 0.93 | *0.98* |
| | μ (s) | -0.06 | -0.01 | *0.00* | -0.07 | 0.00 | *0.00* |
| | σ (s) | 0.24 | *0.09* | 0.11 | *0.08* | 0.18 | 0.09 |
| | MAE (s) | 0.06 | 0.02 | *0.01* | 0.09 | 0.09 | *0.01* |
| S | Recall | 0.76 | *0.99* | *0.99* | 0.94 | 0.61 | *0.95* |
| | Precision | 0.91 | *0.99* | *0.99* | 0.85 | *0.99* | 0.98 |
| | F1 | 0.83 | *0.99* | *0.99* | 0.89 | 0.75 | *0.96* |
| | μ (s) | -0.03 | *-0.00* | -0.03 | -0.09 | -0.03 | *-0.01* |
| | σ (s) | *0.13* | 0.16 | 0.19 | 0.15 | *0.11* | 0.21 |
| | MAE (s) | *0.10* | 0.11 | 0.14 | 0.13 | *0.09* | 0.15 |

### 4.2 Downstream Model 2: Magnitude Estimation Model

Reliable magnitude estimates are crucial for analyzing seismic activity and assessing earthquake risks (Boitz et al., 2024; Firetto Carlino et al., 2022; Gulia and Wiemer, 2019; Herrmann et al., 2022; Taroni and Carafa, 2023). We utilizes MagNet (Mousavi and Beroza, 2020) to assess whether SWaG accurately captures magnitude information from waveforms. MagNet can learn distance-dependent and site-dependent functions directly from the training data. We initially trained MagNet using transfer learning sets, with the magnitude distribution shown in Figure 4a. Subsequently, we trained MagNet using 100,000 waveforms generated by SWaG, as depicted in Figure 4b. Finally, both models were used to predict the tests, and were compared against their magnitude labels (Figure 4c and 4d).

The results indicate that SWaG can effectively create real waveforms based on station and magnitude, as well as epicenter distance derived from S-P arrival times. Notably, the model trained on real waveforms with uneven magnitude distribution tends to overestimate smaller magnitudes while underestimating larger ones, a common challenge in AI-based magnitude estimation models (Mousavi and Beroza, 2020; Saad et al., 2022). Conversely, the model trained on uniformly distributed generated data significantly alleviates this issue.

Despite using a dataset much larger than the real waveforms to train MagNet, the results exhibited similar levels of prediction deviation compared to those trained with real waveforms. We infer that MagNet was designed to be trained with a small local dataset, and its relatively small model structure may not capture all the detailed information related to magnitude in the waveforms. Additionally, the magnitude labels are local magnitudes averaged from multiple stations, which calculated form the maximum amplitude of S-waves. The SWaG preprocessing involved normalization, leading to the loss of this critical feature, which potentially cause the error which affects the accuracy.

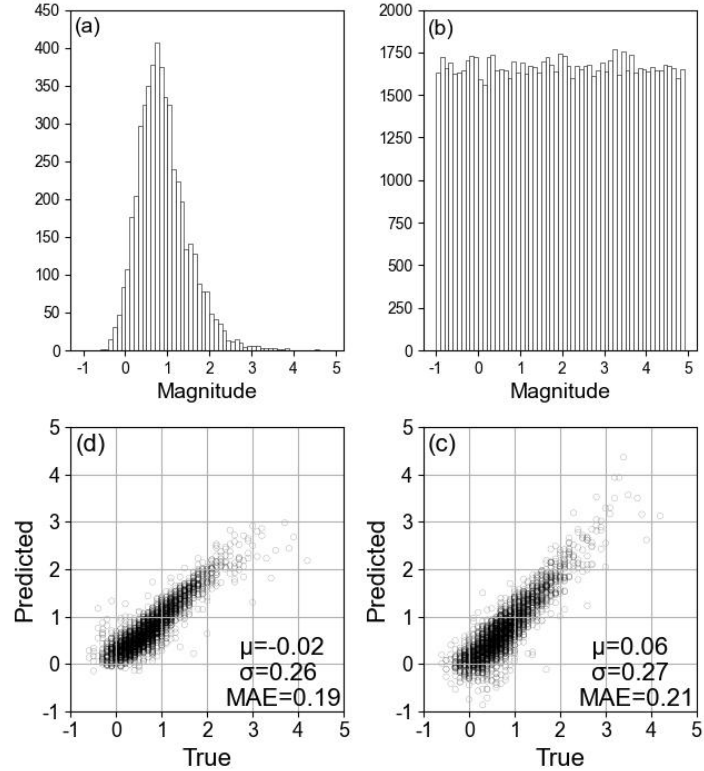

Figure 4. Comparison of MagNet trained with real and generated waveforms. (a) and (b) depict the magnitude distribution of the transfer learning sets and the generated waveform, respectively. (c) and (d) illustrate the comparison of results for models trained with transfer learning sets and generated waveforms, respectively. The horizontal axis represents manually determined magnitudes, and the vertical axis represents model predicted magnitudes.

## 5. Discussion

In evaluating the performance of SWaG, we utilized the transfer learning sets comprising 5,000 waveforms from a single station. However, this posed challenges for stations with low seismicity or limited observation periods. To alleviate this data pressure, we attempted transfer learning using only 500 waveforms, but encountered unsatisfactory results characterized by inaccuracies in arrival times and poor waveform quality. Surprisingly, subsequent experiments with significantly reduced data volumes—500, 200, and even 2 waveforms per station—still generate relatively reliable waveforms in the STEAD dataset. This led us to speculate that overfitting might have influenced training, thereby affecting transfer learning performance with smaller datasets. To address this, we refined our transfer learning approach by adopting a multi-station strategy. Specifically, we utilized data from 10 stations, each containing 500 entries, which notably enhanced the quality of the generated waveforms.

During the phase-picking model test, we observed that bias from models trained on generated waveforms was influenced by the sampling rate of SWaG. We chose a 50Hz sampling rate over 100Hz because SWaG demonstrated poor performance at higher sampling rates (Figure 5). We identified two primary reasons for this observation. First, due to computational constraints, we preprocess data into 500-length tokens using convolution layers before feeding them into DiT Blocks. Thus, advancements in computational power will enhance our ability to effectively process high-sampling-rate waveforms. Secondly, Xu et al. (2019) noted that deep neural networks initially capture dominant low-frequency components swiftly before gradually integrating high-frequency components, which is also pertinent in seismology . Therefore, future generative model designs should consider both high-frequency and low-frequency characteristics, exploring tailored approaches to optimize performance across the board.

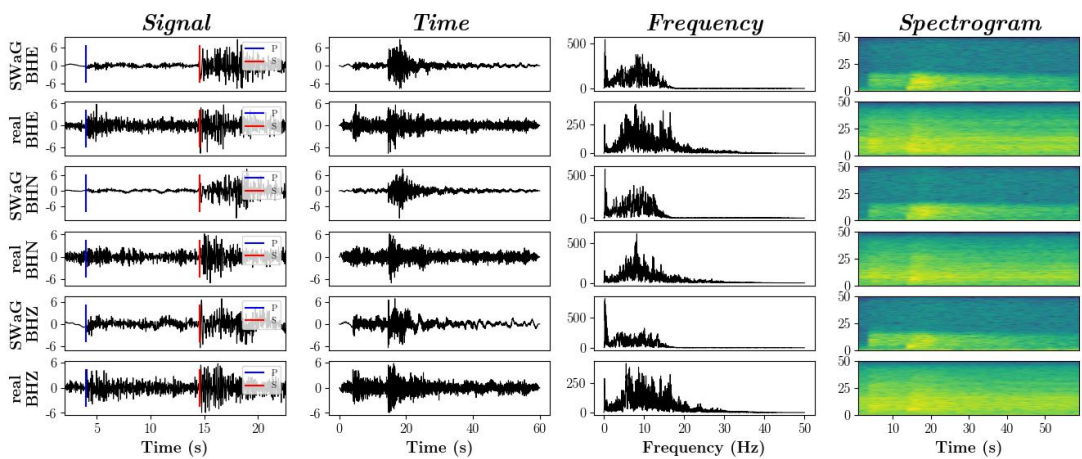

Figure 5. Comparison of generated and real waveforms at sampling rate of 100 Hz. The comparison between the generated waveforms and the real waveforms with the same labels. These comparisons are made across the E, N, and Z components in the signal segments, time domain, frequency domain, and time-frequency domain.

In the results section, we attribute the limited improvement in prediction bias to the characteristics of downstream models and the accuracy of labels. Here, we will delve deeper into the manually determined magnitudes, model predictions trained on real waveforms, and predictions derived from generated waveforms, focusing on magnitude-frequency distributions (referred as label catalog, real waveform-based catalog, and SWaG-based catalog) (Figure 6).

The magnitude-frequency relation were widely observed to follow the Gutenberg–Richter (GR) law globally (Gutenberg and Richter, 1944). Specifically, we employed three methods to estimate the GR $b$-value for each sequence: the widely used maximum likelihood estimation (MLE) (Aki, 1965), the $b$-positive method (Van Der Elst, 2021) derived from differential statistics, and the $K - M$ slope ($KMS$) method (L. Li et al., 2023) based on topological analysis. The MLE and $KMS$ methods were based on magnitude samples, while the $b$-positive method relied on magnitude difference samples (differences between adjacent magnitudes). Considering the incompleteness of earthquake catalogs, only samples above the minimum magnitude of completeness $m_C$ and the minimum complete magnitude difference $m_C'$ were used for parameter estimation. The determination of these thresholds followed the $m_C$ by $b$-value stability method (Cao and Gao, 2002; Woessner, 2005; Li and Luo, 2023). We also presented the evolution of $b$-value estimates under different $m_C$ or $m_C'$ thresholds. If the earthquake sequence followed the GR distribution, we expected the $b$-values obtained by these three methods to be stable above a certain $m_C$ or $m_C'$ threshold and to be consistent across the three methods (Li and Luo, 2023). We found that under the $m_C$ or $m_C'$ determined by the $b$-value stability method, both the SwaG-based catalog and the label catalog adhered quite well to the GR law, with very close $b$-values. For the label catalog, the $b$-values estimated by the three methods are quite stable and consistent above a certain magnitude or magnitude difference cutoff. For the SWaG-based catalog, the $b$-value obtained by the $KMS$ method becomes unstable when $m_C$ is too large, while the other two methods converge to a consistent $b$-value. However, for the real waveform-based catalog, we observed a noticeable deficiency of large earthquakes, with the corresponding $b$-value increasing continuously with the $m_C$ or $m_C'$ cutoff. This feature could be described by the tapered GR distribution (Kagan, 2002), which is thought to reflect the physical limits of large earthquakes. However, we did not believe this was a true representation of the catalog, given the good GR characteristics of the other two catalogs. The results here underscored the potential pitfalls in physical interpretation due to the catalog processing process. Therefore, we infer that the predicted by the model based on the generated waveform may not be the local magnitude, but other magnitudes related to physical quantities such as the duration of the earthquake or the ratio of the maximum amplitude to the minimum amplitude. This magnitude can also conform to the GR relationship and may also characterize the size of the earthquake. In general, the framework can be benefited from choosing appropriate magnitude type, such as using duration magnitude as labels or designing generative models that do not require normalization.

Unlike earlier AI models that depended solely on arrival time (Wang et al., 2021; Chen et al., 2024), SWaG can generate waveforms based on earthquake magnitude. Models trained on these generated waveforms demonstrate the capability to accurately predict the magnitudes of actual seismic event. The constraints associated with arrival time are related to the amplitude and period characteristics of seismic waveforms, which are theoretically independent of regional variations. Consequently, the seismic phase picking model, when trained on regional data, exhibits a degree of generalizability in non-training areas, as evidenced in various seismological studies (Duan et al., 2023, 2024; Feng et al., 2024; Zhang et al., 2022). Inversion of source parameters, such as earthquake magnitude, focal mechanism, and stress drop, is associated with paths and receivers. Models trained within a specific region often lack the ability to generalize effectively, and the limited observational data available in that region may be insufficient to support the training of large-parameter AI models. This inadequacy presents significant challenges to the advancement of AI in seismology (Bergen et al., 2019; Anikiev et al., 2023; Kubo et al., 2024). The establishment of accurate global seismic source parameter datasets could empower future diffusion-based generative models to effectively address these challenges. Furthermore, the potential applications of generative models extend beyond this context. Seismic wave observations are frequently time-consuming, costly, and sparse, and interruptions due to equipment malfunctions can impede seismological research. Generative models hold promise for filling observational gaps, thereby significantly advancing the field of seismology.

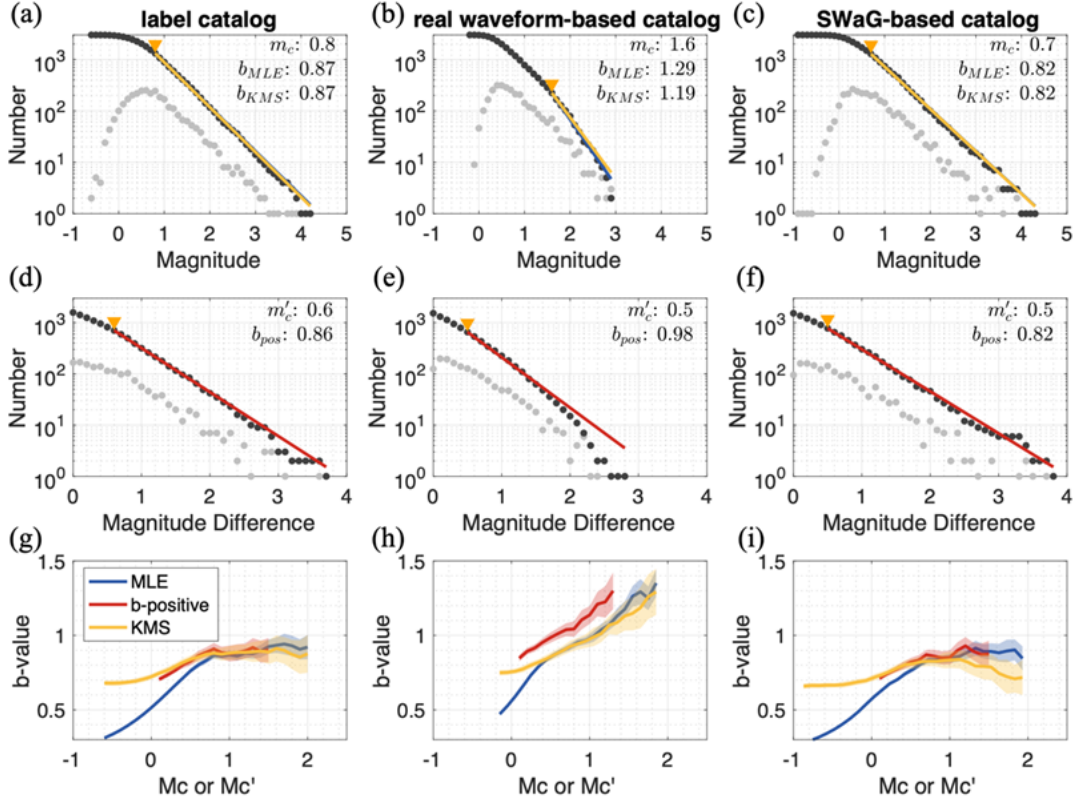

Figure 6. Comparison of magnitude–frequency distribution of different catalog. (a) The magnitude–frequency distribution of the label catalog. The gray circles represent the number of samples equal to a particular magnitude. The black circles represent the number of samples greater than or equal to a particular magnitude. The blue and yellow lines are the optimal Gutenberg–Richter model by the maximum likelihood estimation and the $K - M$ slope method, respectively. (b) The magnitude difference–frequency distribution of the label catalog. The red line is optimal Gutenberg–Richter model by the $b$-positive method. (c) The estimated $b$-values under different minimum magnitude of completeness $m_c$ or the minimum complete magnitude difference $m_c'$. The shaded area is the one standard deviation obtained by the bootstrap methods. (d, e, f) and (h, i, j) Similar to (a, b, c), but for real waveform-based catalog and SwaG-based catalog.

## 6. Conclusion

This study introduces a seismic wave generator (SWaG) built on a multi-condition diffusion model with a transformer architecture. The model can be fine-tuned using a small amount of local data through transfer learning, enabling the creation of numerous region-specific waveforms that adhere to specified conditions. Phase picking models trained on these generated waveforms outperform standard models trained on conventional datasets, achieving higher precision and recall when predicting real, manually annotated waveforms. This demonstrates SWaG can effectively address the generalization challenges faced by phase picking models. Additionally, magnitude estimation models trained on generated waveforms can reduce biases in predicting real event magnitudes caused by uneven training data distribution. These findings suggest that generated waveforms could support regional earthquake magnitude predictions and indicate that future diffusion-based generative models may address critical challenges in AI determining seismic source parameters.

## Acknowledgments

This study was funded by the National Key Research and Development Program of China (2021YFC3000703), the Central Public-interest Scientific Institution Basal Research Fund (CEAIEF20240401), the National Natural Science Foundation of China (42174066) and supported by the Shanghai Artificial Intelligence Laboratory. We thank Linxuan Li at California Institute of Technology for his help in the task of magnitude estimation.

## Data Availability

The STEAD dataset is available at: https://github.com/smousavi05/STEAD . The code and model of PhaseNet is available at https://github.com/AI4EPS/PhaseNet. The code and model of EQTransformer is available at https://github.com/smousavi05/EQTransformer. The code of MagNet is available at https://github.com/smousavi05/MagNet. The code of Seisbench is available at https://github.com/seisbench/seisbench. The data, code, and model of SWaG is available at https://github.com/Lonngfei/SWaG.